\documentclass[conference]{IEEEtran}
\IEEEoverridecommandlockouts
\usepackage{cite}
\usepackage{amsmath,amssymb,amsfonts}
\usepackage{algorithmic}
\usepackage{graphicx}
\graphicspath{{./img/}}
\DeclareGraphicsExtensions{.pdf,.jpeg,.png}
\usepackage{textcomp}
\usepackage{xcolor}
\def\BibTeX{{\rm B\kern-.05em{\sc i\kern-.025em b}\kern-.08em
    T\kern-.1667em\lower.7ex\hbox{E}\kern-.125emX}}

\usepackage[para]{footmisc}
\usepackage{url}
\usepackage{nohyperref} 
\usepackage[nohyperlinks,nolist]{acronym} 
\usepackage{booktabs}
\usepackage{tabularx}
\newcommand*{\MinNumber}{0.0}%
\newcommand*{\MidNumber}{0.5}%
\newcommand*{\MaxNumber}{1.0}%
\newcommand{\ApplyGradient}[1]{%
        \ifdim #1 pt > \MidNumber pt
            \pgfmathsetmacro{\PercentColor}{max(min(100.0*(#1 - \MidNumber)/(\MaxNumber-\MidNumber),100.0),0.00)} %
            \hspace{-0.33em}\colorbox{green!\PercentColor!yellow}{#1}
        \else
            \pgfmathsetmacro{\PercentColor}{max(min(100.0*(\MidNumber - #1)/(\MidNumber-\MinNumber),100.0),0.00)} %
            \hspace{-0.33em}\colorbox{red!\PercentColor!yellow}{#1}
        \fi
}
\newcolumntype{Y}{>{\centering\arraybackslash}X}
\newcolumntype{L}{>{\arraybackslash}X}
\newcolumntype{R}{>{\raggedleft\arraybackslash}X}
\newcolumntype{C}[1]{>{\centering\arraybackslash}p{#1}}
\newcolumntype{G}[1]{>{\collectcell\ApplyGradient}#1<{\endcollectcell}}
\usepackage{multirow}
\usepackage[per-mode=symbol]{siunitx}
\usepackage{pgfplots}
\pgfplotsset{compat=newest}
\usepgfplotslibrary{units}
\makeatletter
\pgfplotsset{
 unit code/.code 2 args=
   \begingroup
   \protected@edef\x{\endgroup\si{#2}}\x
}
\pgfplotsset{
  /pgfplots/ybar legend/.style={
    legend image code/.code={
      \draw [#1] (0cm,-0.1cm) rectangle (0.15cm,0.2cm);
    };
  }
}
\pgfplotsset{
  /pgfplots/xbar legend/.style={
    legend image code/.code={
      \draw [#1] (-0.1cm,-0.05cm) rectangle (0.2cm,0.1cm);
    };
  }
}
\makeatother
\usetikzlibrary{pgfplots.groupplots}
\usetikzlibrary{matrix}
\usetikzlibrary{patterns}
\usepackage[absolute,showboxes]{textpos}

\definecolor{corered}{RGB}{189, 40, 22}
\definecolor{coregray}{RGB}{191, 191, 191}
\definecolor{coredarkgray}{RGB}{51,51,51}
\definecolor{coreblue}{RGB}{1, 49, 121}
\definecolor{coregreen}{RGB}{121, 170, 144}
\definecolor{plotred}{RGB}{215,25,28}
\definecolor{plotorange}{RGB}{253,174,97}
\definecolor{plotyellow}{RGB}{255,255,191}
\definecolor{plotgreen}{RGB}{171,221,164}
\definecolor{plotblue}{RGB}{43,131,186}

\begin{document}

\bstctlcite{my:BSTcontrol}

\title{A Framework for the Systematic Assessment of Anomaly Detectors in Time-Sensitive\\ Automotive Networks
}

\author{\IEEEauthorblockN{Philipp Meyer, Timo H\"ackel, Teresa L\"ubeck, Franz Korf, and Thomas C. Schmidt}\IEEEauthorblockA{\href{http://www.haw-hamburg.de/ti-i}{\textit{Dept. Computer Science}},
\href{http://www.haw-hamburg.de/ti-i}{\textit{Hamburg University of Applied Sciences}}, Germany \\
\{\href{mailto:philipp.meyer@haw-hamburg.de}{philipp.meyer}, \href{mailto:timo.haeckel@haw-hamburg.de}{timo.haeckel}, \href{mailto:teresa.luebeck@haw-hamburg.de}{teresa.luebeck}, \href{mailto:franz.korf@haw-hamburg.de}{franz.korf}, \href{mailto:t.schmidt@haw-hamburg.de}{t.schmidt}\}@haw-hamburg.de}
}

\maketitle

\setlength{\TPHorizModule}{\paperwidth}
\setlength{\TPVertModule}{\paperheight}
\TPMargin{5pt}
\begin{textblock}{0.8}(0.1,0.02)
     \noindent
     \footnotesize
     If you cite this paper, please use the original reference:
     Philipp Meyer, Timo H\"ackel, Teresa L\"ubeck, Franz Korf, and Thomas C. Schmidt. "A Framework for the Systematic Assessment of Anomaly Detectors in Time-Sensitive Automotive Networks," In: \emph{Proceedings of the 15th IEEE Vehicular Networking Conference (VNC)}. IEEE Press, May 2024.
\end{textblock}


\begin{abstract}
	Connected cars are susceptible to cyberattacks. Security and safety of future vehicles highly depend on a holistic protection of automotive components, of which the time-sensitive backbone network takes a significant role.
  These onboard \acp{TSN} require  monitoring for safety and -- as versatile platforms to host \acp{NADS} -- for security. Still a thorough evaluation of anomaly detection methods in the context of hard real-time operations, automotive protocol stacks, and domain specific attack vectors is missing along with appropriate input datasets.
	In this paper, we present an assessment framework that allows for reproducible, comparable, and rapid evaluation of detection algorithms. It is based on a simulation toolchain, which contributes configurable topologies, traffic streams, anomalies, attacks, and detectors.
  We demonstrate the assessment of \acp{NADS} in a comprehensive in-vehicular network with its communication flows, on which we model traffic anomalies.
  We evaluate exemplary detection mechanisms and reveal how the detection performance is influenced by different combinations of \ac{TSN} traffic flows and anomaly types.
  Our approach translates to other real-time Ethernet domains, such as industrial facilities, airplanes, and UAVs.
\end{abstract}


\begin{IEEEkeywords}
  Automotive, Security, Anomaly Detection, Time-Sensitive Networks, Simulation
\end{IEEEkeywords}

\begin{acronym}
	\acro{ACDC}[ACDC]{Automotive Cyber Defense Center}
	\acro{AD}[AD]{Anomaly Detection}
	\acro{ADAS}[ADAS]{Advanced Driver Assistance System}
	\acro{ADS}[ADS]{Anomaly Detection System}
	\acroplural{ADS}[ADSs]{Anomaly Detection Systems}
	\acro{API}[API]{Application Programming Interface}
	\acro{AUTOSAR}[AUTOSAR]{AUTomotive Open System ARchitecture}
	\acro{AVB}[AVB]{Audio Video Bridging}
	\acro{ARP}[ARP]{Address Resolution Protocol}
	\acro{ATS}[ATS]{Asynchronous Traffic Shaping}
	\acro{BE}[BE]{Best-Effort}
	\acro{CAN}[CAN]{Controller Area Network}
	\acro{CBM}[CBM]{Credit Based Metering}
	\acro{CBS}[CBS]{Credit Based Shaping}
	\acro{CMI}[CMI]{Class Measurement Interval}
	\acro{CoRE}[CoRE]{Communication over Realtime Ethernet}
	\acro{CT}[CT]{Cross Traffic}
	\acro{DoS}[DoS]{Denial of Service}
	\acro{DDoS}[DDoS]{Distributed \acl{DoS}}
	\acro{DPI}[DPI]{Deep Packet Inspection}
	\acro{ECU}[ECU]{Electronic Control Unit}
	\acroplural{ECU}[ECUs]{Electronic Control Units}
	\acro{ESC}[ESC]{Electronic Stability Control}
	\acro{FN}[FN]{False Negative}
	\acro{FP}[FP]{False Positive}
	\acro{HBOS}[HBOS]{Histogram-based Outlier Score}
	\acro{HTTP}[HTTP]{Hypertext Transfer Protocol}
	\acro{HMI}[HMI]{Human-Machine Interface}
	\acro{IA}[IA]{Industrial Automation}
	\acro{IAM}[IAM]{Identity- and Access Management}
	\acro{ICT}[ICT]{Information and Communication Technology}
	\acro{IDS}[IDS]{Intrusion Detection System}
	\acroplural{IDS}[IDSs]{Intrusion Detection Systems}
	\acro{IEEE}[IEEE]{Institute of Electrical and Electronics Engineers}
	\acro{IoT}[IoT]{Internet of Things}
	\acro{IP}[IP]{Internet Protocol}
	\acro{IVN}[IVN]{In-Vehicle Network}
	\acroplural{IVN}[IVNs]{In-Vehicle Networks}
	\acro{LIN}[LIN]{Local Interconnect Network}
	\acro{MTU}[MTU]{Maximum Transmission Unit}
	\acro{MOST}[MOST]{Media Oriented System Transport}
	\acro{NAD}[NAD]{Network Anomaly Detection}
	\acro{NADS}[NADS]{Network Anomaly Detection System}
	\acroplural{NADS}[NADSs]{Network Anomaly Detection Systems}
	\acro{NID}[NID]{Network Intrusion Detection}
	\acro{NIDS}[NIDS]{Network Intrusion Detection System}
	\acroplural{NIDS}[NIDSs]{Network Intrusion Detection Systems}
	\acro{OEM}[OEM]{Original Equipment Manufacturer}
	\acro{PCAP}[PCAP]{Packet Capture}
	\acro{PCAPNG}[PCAPNG]{PCAP Next Generation Dump File Format}
	\acro{PCP}[PCP]{Priority Code Point}
	\acro{PSFP}[PSFP]{per-stream filtering and policing}
	\acro{RC}[RC]{Rate-Constrained}
	\acro{REST}[ReST]{Representational State Transfer}
	\acro{SDN}[SDN]{Software-Defined Networking}
	\acro{SOA}[SOA]{Service-Oriented Architecture}
	\acroplural{SOA}[SOAs]{Service-Oriented Architectures}
	\acro{SOME/IP}[SOME/IP]{Scalable service-Oriented MiddlewarE over IP}
	\acro{SR}[SR]{Stream Reservation}
	\acro{SRP}[SRP]{Stream Reservation Protocol}
	\acro{SVM}[SVM]{Support Vector Machine}
	\acro{TAS}[TAS]{Time-Aware Shaping}
	\acro{TCP}[TCP]{Transmission Control Protocol}
	\acro{TDMA}[TDMA]{Time Division Multiple Access}
	\acro{TN}[TN]{True Negative}
	\acro{TP}[TP]{True Positive}
	\acro{TSN}[TSN]{Time-Sensitive Networking}
	\acroplural{TSN}[TSNs]{Time-Sensitive Networks}
	\acro{TSSDN}[TSSDN]{Time-Sensitive Software-Defined Networking}
	\acro{TT}[TT]{Time-Triggered}
	\acro{TTE}[TTE]{Time-Triggered Ethernet}
	\acro{UDP}[UDP]{User Datagram Protocol}
	\acro{QoS}[QoS]{Quality-of-Service}
	\acro{V2X}[V2X]{Vehicle-to-X}
	\acro{WS}[WS]{Web Services}
	\acro{ZC}[ZC]{Zone-Controller}
	\acroplural{ZC}[ZCs]{Zone-Controllers}
\end{acronym}

\section{Introduction}
\label{sec:introduction}

Future vehicle functions for assisted and autonomous driving rely on interconnected safety critical components and voluminous real-time streams from sensors such as cameras or LIDARs.
In consequence, internal automotive networks are required to concurrently cope with significantly higher bandwidth and differentiated time-critical communication across domains.
This makes the \ac{IVN} a critical component, which needs firm, multi-sided protection---in particular in the presence of global connectivity.
Automotive Ethernet is the emerging candidate for future in-car networks and---augmented by the IEEE 802.1 \ac{TSN} standard~\cite{ieee8021q-22}---will provide a solution for distinct \ac{QoS} guarantees in heterogeneous automotive settings.

Already today, cars span a large attack surface comprising physical and wireless interfaces~\cite{cmkas-ceaas-11}.
Attacks on the in-car network can lead to severe incidents such as disruptive operation, loss of control, or ramming attacks.
Attacks on the \ac{IVN} can lead to anything from operational disruptions and loss of control to safety-critical accidents.
Advanced network security is essential for the safety of automotive functions.
Network monitoring is a key component for network protection~\cite{wmslk-sdvtm-17,mbzls-ssaJR-18}.
Monitoring in-car communication enables the detection and mitigation of errors or attacks.

Evaluating the performance of a \ac{NADS} depends on the availability of suitable data.
Existing datasets, like CIC-IDS~2017~\cite{shg-gniJR-18l}, are of limited use because they lack domain-specific communication protocols, patterns, and anomalies.

An evaluation framework for various \acp{NADS} not only provides insights into the dependability in operation, but also promises reproducibility, comparability, and rapid assessment.

In this work, we present a comprehensive assessment framework for \acp{NADS}, which we release as open-source software, and its exemplary application to the automotive domain.
Our contributions cover two key areas:
\begin{itemize}
	\item {\em Generating datasets:}
	We utilize a comprehensive simulation environment with individual configurations, automatic labeling, and replaceable attack models.
	A specific configuration can describe a multitude of scenarios containing benign and abnormal communication.
	Networks can be created based on artificial or real communication specifications.
	The resulting dataset library consists of labeled \ac{PCAP} files.
\item {\em Evaluating \ac{NADS}:}
	Our \ac{NADS} framework allows evaluating and comparing different combinations of traffic filters, metrics, and algorithms.
	High modularity enables rapid tests of various compositions using identical datasets.
	The \ac{PCAP} format also facilitates the use of recorded datasets from real-world deployments.
\end{itemize}
Together, we contribute a toolchain that enables reproducible and comparable assessments for \ac{NADS}.
In our case study, we demonstrate the application of the toolchain in automotive networks.
The toolchain can be applied to other domains since all components are expandable and the domain-specific aspects mainly depend on configurations.

The remainder of this paper is structured as follows.
Section~\ref{sec:background} gives an overview on background and related work.
We present the framework for assessing \acp{NADS} performance in \ac{TSN} networks in Section~\ref{sec:toolchain}, which consists of scenario configurations, simulation environment, dataset generation, and a \ac{NADS} composition under evaluation.
Section \ref{sec:case_study} guides through a case study, in which the framework enfolds its impact, enabling reproducible \ac{NADS} performance evaluations different scenarios of an \ac{IVN}.
Finally, Section \ref{sec:conclusion} concludes the paper with an outlook on future work.

\section{Background and Related Work}
\label{sec:background}

Traditional in-car networks subsume a set of \acp{ECU} linked to one of multiple \ac{CAN} busses which are interconnected via one central gateway. 
Each bus represents a domain of the vehicle (e.g., motor control, infotainment).
A single flat Ethernet network in a zonal topology is promising candidate for future in-car architectures~\cite{brkw-aeajr-17}.
The industry standard AUTOSAR defines the automotive Ethernet stack to be a \acl{SOA} consisting of TSN, IP, UDP, and e.g., SOME/IP.
In the transition phase, legacy components must be linked via gateways to the \ac{TSN} backbone of the car.
In our case study, a \ac{TSN} backbone in the transition phase is used.

\ac{TSN}~\cite{ieee8021q-22} is a collection of standards for \ac{QoS} in Ethernet networks.
\ac{TSN} organizes packets into streams.
Each stream can be assigned to different traffic classes.
Synchronous (e.g., \ac{TAS}) and asynchronous (e.g., \ac{ATS}, \ac{CBS}) traffic shaping techniques can be used concurrently on the same links.
Individual shaping configurations impose different \ac{QoS} guarantees and specific patterns on traffic behavior.

Simulation is widely used to evaluate \ac{TSN} topology designs, traffic shaping configurations, and protocol combinations~\cite{hgo-titJR-16l,jlhxw-tsnJR-18l,fhcnd-nsiJR-19l}.
In previous work, we built a simulation environment for evaluation of \ac{IVN} designs~\cite{mkss-smcin-19}.
OMNeT++\footnote{\url{omnetpp.org}} and INET\footnote{\url{inet.omnetpp.org}} provide a toolbox for network simulation (e.g., Ethernet, IP, UDP).
Recent additions are \ac{TSN} features,
such as clock models, time synchronization, frame replication, and traffic shaping.
This comprehensive collection allows extensible and configurable simulation of complex \ac{TSN} networks.
We utilize this framework to build realistic networks, traffic, and anomalies to generate labeled traffic captures.

Security and privacy is a major concern of future \acp{IVN}~\cite{rgks-oasJR-20l}.
Traditional in-car infrastructures are closed systems with limited access to the outside world.
Accordingly, \ac{IVN} security has not been considered in former automotive engineering leaving major vulnerabilities in existing vehicles~\cite{mv-reupv-15}.
Interfaces of modern cars span a large attack surface while offering access from the outside world~\cite{cmkas-ceaas-11}.
Cars are vulnerable to cyber-attacks~\cite{dzjal-adiJR-20,pshf-tamJR-21}.
Attacks on interfaces can spread into the \ac{IVN} and vice versa.
Every \ac{IVN} component is a potential victim of compromise~\cite{dzjal-adiJR-20}.
New regulations demand advanced protective measures such as updates and monitoring for entire future car lifecycles~\cite{unece-wp29,iso-sae-21434}.
In previous work, we integrated a versatile security architecture into a production car~\cite{mhlsd-dsivi-20}.

Intrusion detection of malicious behavior is key to limit the impact of cyber-attacks~\cite{wmslk-sdvtm-17} because it enables execution of countermeasures that prevent further damage.
Anomaly detection is a behavioral approach to intrusion detection which spots misbehavior by identifying deviations from normal behavior.
Thus, it enables the possible detection of zero-day attacks.
The implementation of \acp{NADS} can differ widely in input metrics and detection methods~\cite{bbk-nadms-14,rmwh-sadJR-18,hkk-tiiJR-22l}.
Detection performance can be significantly influenced by the combination of metrics, algorithms, and individual network traffic patterns.
Assessment of \acp{NADS} in real-world deployments can be done by manual inspection of detection results~\cite{hgsd-amaJR-19} in late phases of a vehicle development cycle.
Systematic assessment methods based on relevant and comparable datasets enable early, fast, and low-effort evaluation.
Further, identification of candidates ahead of deployment is possible.

Several datasets tailored for evaluation of intrusion detection exist~\cite{shg-gniJR-18l,tl-raiJR-20l,lckll-cadJR-23l}.
Those datasets, however, must match the deployment environment of the \ac{NADS} to get meaningful results.
A major challenge is the realistic representation of attacks to which a vehicle could fall victim~\cite{sgkks-ccsJR-24l}.
Often datasets lack specific features of the target environment (e.g., physical layer, protocols, attacks, metrics).
The project PIVOT\footnote{\url{pivot-auto.org}} 
aim to support vehicle security research by providing a valuable platform where among other things datasets can be shared. 
Existing datasets are unable to cover specific cases.
Customizable generation enable datasets to be tailored to specific cases~\cite{shg-gniJR-18l,rsw-idhJR-18l}.
\ac{TSN} based networks impose specific patterns on traffic behavior.
Further, automotive attack vectors are partially domain-specific~\cite{lflhs-adgJR-22l}.
A dataset generator needs to support those particularities to be applicable.
We use a network simulation framework that is highly customizable and allows the integration of external datasets as stimuli.

Simulations allow for fast and flexible generation of domain-specific data~\cite{ispt-nnbJR-21l}.
On the other hand, results are only as good as the model in use.
The past has shown that synthetic data can differ significantly from real-world results~\cite{m-tidJR-00l}.

The open-source attack simulation framework Simutack~\cite{fmlhs-sasJR-23l} implements a comprehensive model of automotive systems that can cover realistic attack scenarios.
The holistic approach of Simutack simulates attacks within the scope of the entire vehicle with its subsystems and its environment.
A wide range of attacks can be tested in a wide variety of traffic situations.
Simutack shows that simulation, despite its disadvantages, is needed to enable safe and low-cost creation of realistic datasets for intrusion detection.
Focus of our framework is the generation of \ac{IVN} datasets and targets the assessment of network-based intrusion detection.
To achieve this, we use a simulation environment where network technologies and in particular \ac{TSN} are covered by comprehensive implementations of protocols and traffic shaping techniques.

Like Simutack, we use an open-source and community-driven simulation environment that enables early generation of datasets while providing full transparency into implementation, uncovered issues, and limitations.
In contrast, our simulation operates on the system level where it supports variable topologies and highly customizable scenarios.
This enables a detailed modelling of traffic patterns and the exploration of domain-specific anomalies.
In this work, we consider a generic attack model that is based on fundamental classes of link layer anomalies (e.g., injection, elimination, reordering).
The behavior of higher layer attacks consists of a combination of those anomaly classes. 
Our framework supports integration of further attack models and packet captures.

\section{Tools for Assessing Network Anomaly Detectors}
\label{sec:toolchain}

All tools are integrated into a toolchain which enables assessment of network anomaly detectors (s. Fig.~\ref{fig:toolchain}).
All contributions of this work including case study configurations are published open source and are available on GitHub\footnote{\url{github.com/CoRE-RG/NADS-Assessment-Tools}}.
In addition of providing required components, the existing tools also simplify the integration of further configurations, attack models, datasets, and detectors.
Real-car inputs like \ac{IVN} configurations and recorded traces can be interlaced.
The toolchain consists of four main components: network scenario, simulation environment, dataset library, and \ac{NADS}.
A network scenario provides the parameters that are configured for a specific simulation.
By simulating scenarios, labeled network traces are generated that yield a dataset library.
Together with the \ac{NADS} framework a detector configuration can be tested.

\subsection{In-Car Network Scenario}
\label{subsec:simulation}
Scenario configurations are domain-specific definitions.
This component defines an individual network topology, baseline traffic, protocol stack, and scenarios.
The following configurations must be defined for a complete scenario.

\subsubsection{Topology}
A topology is a specification of the network architecture with device types (e.g., \acp{ECU} and switches), names, position, and Ethernet links.
Real in-car networks can be modelled precisely in simulation.
A finished topology is a complete network architecture definition.

\begin{figure}
	\centering
	\includegraphics[width=1\columnwidth, trim= 0.6cm 0.6cm 0.6cm 8.6cm, clip=true]{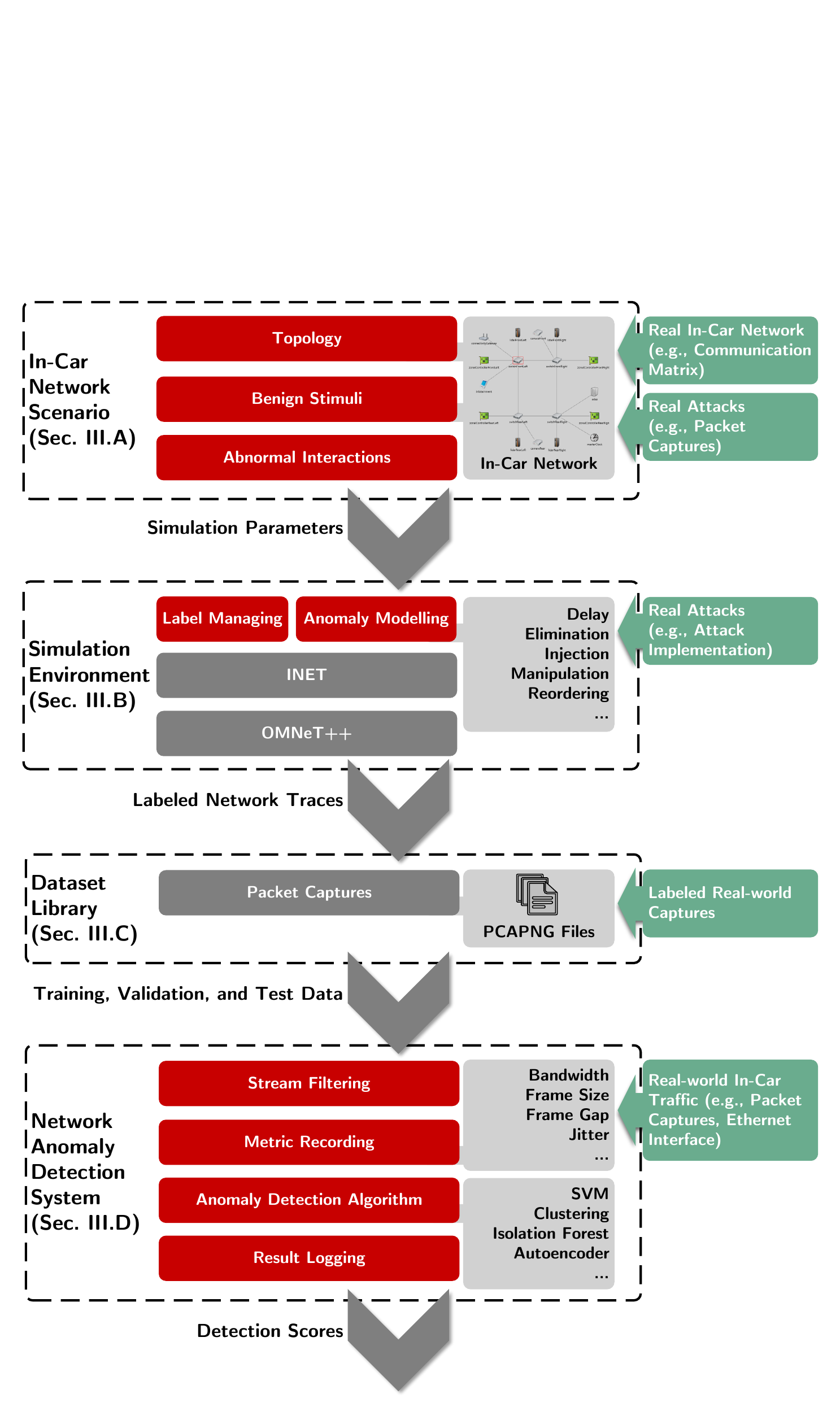}
	\vspace{-20pt}
	\caption{Toolchain for assessment of \acp{NADS}. Red are contributions of this work. Dark gray represent existing tools. Light gray depicts implementation examples. Green indicates optional input from empirical measurements.}
	\label{fig:toolchain}
	\vspace{-6pt}
\end{figure}

\subsubsection{Benign Stimuli}
The size, timing and variation of produced messages must be defined or derived from existing \ac{IVN} definitions.
Stimuli definitions can be derived from real traffic patterns like application traces and OEM communication matrices. 
Existing traffic captures can additionally be brought in.
Stimuli composition also contains definition of packet sources and sinks, the \ac{TSN} configurations for all used protocols, and resulting streams with their schedules.
For example, an application that sends through a UDP socket needs to be mapped to a \ac{TSN} stream multicast MAC address and Q-Tag with VLAN ID and \ac{PCP}.
For each hop possible policing, shaping, and replication modules must be configured to match the requirements of the stream.

\subsubsection{Abnormal Interactions}
\label{subsubsec:abnormal_interactions}
This module represents the configuration of the attack model.
Interaction with existing packets (e.g., elimination, delay, manipulation) or further stimuli (e.g., additional packet generators) integrate anomalies.
In addition, real attack traces can be used as a stimuli input for simulation.

\subsection{Simulation Environment}
\label{subsec:simulation_env}
Detailed link layer simulation is essential to create realistic network traces.
The used environment is based on the discrete event simulator OMNeT++ which is widely used for simulation of networks.
On top of that we use the INET framework which provides comprehensive networking modules for \ac{TSN} (e.g., gPTP, Enhancements for Scheduled Traffic, Frame Replication and Elimination for Reliability) and higher layers (e.g., IP, UDP, TCP).
We extend the environment with labeling support and anomaly models.
Multiple variations of a simulation are defined quickly with inheritance of existing configurations.
It is possible to record incoming and outgoing packets per network interface on each device.
Because we focus on the \ac{IVN}, we do not simulate the whole car (like e.g., CARLA) or interactions between vehicles (like e.g., SUMO).

\subsubsection{Label Managing}
\label{subsubsec:label_managing}
Label tagging for two simultaneous labels is implemented.
One for phases and one for individual packets.
A phase label is set for all packets in the period where a named phase is active.
Packet labels are individually set for packets that are subject to an anomaly.
Other packets are labeled benign (BENIGN).
Missing packets are labeled indirectly with a special benign label (BENIGN RECOVERED) tagged to the first normal packet after every anomaly.

\subsubsection{Anomaly Modelling}
The framework currently contains five elemental link layer anomalies: Delay, elimination, injection, manipulation, and reordering.
The models support automatic labeling of affected packets and can be used on different layers in the protocol stack.
Those anomaly components represent a generic attack model that can be exchanged or combined.
Real-world attacks can be implemented with suitable interactions on selected layers.

\subsection{Dataset Library}
\label{subsec:dataset_lib}
With multiple simulations, where each simulation generates \ac{PCAP} files, a dataset library is created.
It consists of labeled \ac{PCAPNG} files that contain all traffic traversing selected interfaces input and/or output.
The labels (cf. Sec.~\ref{subsubsec:label_managing}) are placed as packet comments in the \ac{PCAPNG} files. 
The dataset library decouples the simulation from the \ac{NADS} evaluation.
Labeled real-world traces can be integrated and any \ac{NADS} implementation that is able to process the contained \ac{PCAP} files can use the library.

\begin{figure}
	\centering
	\includegraphics[width=1\columnwidth, trim= 0cm 0cm 0cm 0cm, clip=true]{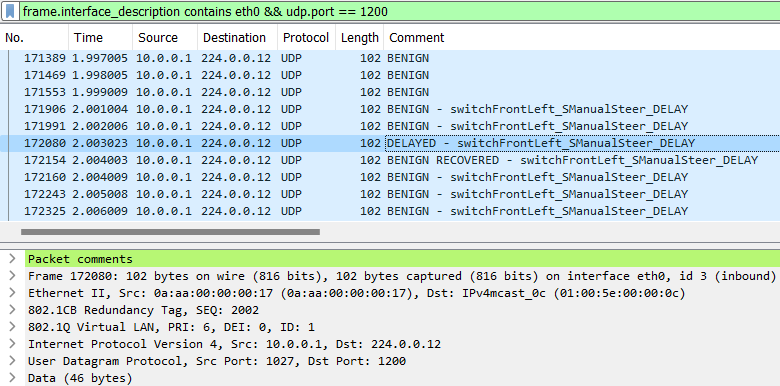}
	\caption{\ac{PCAPNG} excerpt of a stream (UDP destination port 1200) with a starting abnormal scenario observed on a switch interface (eth0).}
	\label{fig:wireshark}
\end{figure}

Figure \ref{fig:wireshark} shows an excerpt of a labeled \ac{PCAPNG} capture recorded in simulation.
The display filter is set to show only steer-by-wire communication packets (UDP destination port 1200) traversing on network interface eth0.
The packet and phase label in the packet comment is separated by a dash.
Until \SI{2}{\second} the packet labels are normal (BENIGN) and the phase label is empty.
At a time of \SI{2}{\second} the phase label is set to mark an abnormal phase.
This scenario delays packets by \SI{10}{\micro\second} with a probability of \SI{50}{\percent} and minimum distance between abnormal actions of \SI{10}{\milli\second}.
The first delayed packet is marked with a packet label (DELAYED) at \SI{2.003023}{\second}.
The next normal packet is marked to be the first normal packet after an abnormal packet (BENIGN RECOVERED).
This allows labeling of anomalies that arise due to absence of packets.

Multiple packet captures for baseline and abnormal scenarios compose the dataset library.
It is extendable though additional simulations or real-world datasets which are labeled accordingly.
Such a dataset library can be split to create training, validation, and testing sets for evaluation of \acp{NADS}.

\subsection{Network Anomaly Detection System}
\label{subsec:nads}
Any \ac{NADS} that is able to process the dataset format can be evaluated with the given library.
Our \ac{NADS} framework is a Python-based implementation of a processing pipeline that operates with four interchangeable layers (cf. Fig. \ref{fig:toolchain}).

\subsubsection{Stream Filtering}
The filtering consists of interchangeable input sources that enable filtering of individual streams from real interfaces (raw socket, Scapy\footnote{\url{scapy.net}}, PyShark\footnote{\url{kiminewt.github.io/pyshark}}), simulation (TCP socket), or \ac{PCAPNG} files (PyShark).
Dependent on input, the program time is synchronized on system clock, simulation time, or packet timestamps.
This interface enables the use of the same implementation and configuration in different environments and allows the comparison of dataset assessments with real-world results.

\subsubsection{Metric Recording}
Feature extraction is represented as a selectable combination of metrics.
It can record single or a set of metrics in selected intervals.
Implemented metrics are bandwidth, frame size, frame gap, and cycle jitter.
Further metrics can be added to the layer.
It enables evaluation of metric combinations influence on detection performance.

\subsubsection{Anomaly Detection Algorithm}
Training and detection tasks can be executed by interchangeable algorithms.
Provided metrics are used as input for any algorithm.
Implemented algorithms build upon widely used frameworks like scikit-learn\footnote{\url{scikit-learn.org}} (e.g., isolation forest, one class \ac{SVM}, k-Means, mean-shift) and TensorFlow\footnote{\url{www.tensorflow.org}} (e.g., Autoencoder).
Again, further algorithms can be added with their dependencies.
Interchangeable algorithms allow the comparison of their performance under repeatable and identical conditions.

\subsubsection{Result Logging}
The last part logs the detection algorithm results.
Result logging generates general statistics, a confusion matrix and calculates performance metrics such as precision and recall.
With the resulting detection scores this modular design enables comprehensive evaluations of \ac{NADS} approaches and their performance with varying input on different traffic pattern.

\subsection{Capabilities, Opportunities and Limitations}
\label{subsec:bottomline}
The presented toolchain opens a multidimensional space of configuration options for generating datasets (topologies, protocol stacks, benign stimuli, abnormal interactions/attack models) and \ac{NADS} composition (input sources, stream filters, feature extraction metrics, detection algorithms).
All options are extendable and can be adapted to new use cases.
Single options can be changed while others remain constant to evaluate their influence on the detection performance.

This multidimensional space enables a systematic and reproducible approach to the assessment of \acp{NADS} under various comparable conditions.
Through vast parameter studies a large matrix of approaches can be evaluated to discover promising candidates.
This is a major milestone for future work.

However, it must be noted that real-world tests are not replaced by this toolchain.
Exchange with real world results is necessary at various points to keep the quality of the results high.
The toolchain allows interfacing with various real-world data at different layers.
Simulation statistics need to be validated with analytical expectations and real world data.
This ensures that generated packet captures are close to reality.
The \ac{NADS} implementation allows the input of real-world data through recorded \ac{PCAPNG} files and by using real network interfaces.
This acts as a corrective for results derived from simulated datasets and can provide indications of real-time feasibility of anomaly detection approaches.
Discovered promising combinations must be tested in the real world to ensure their applicability.

\section{Assessing Automotive Network Anomaly Detectors: A Case Study}
\label{sec:case_study}

In this case study, we adjust three configuration option dimensions (abnormal interactions, stream filters, and detection algorithms).
Other dimensions (topology, protocol stack, benign stimuli, extracted metrics) are kept constant.
With this, we present three exemplary scenarios for assessing automotive \ac{NADS} performance, demonstrating the flexibility of our framework to evaluate different scenarios and algorithms in fast cycles.
Our selection represents a small subset of typical misbehavior including packet loss, protocol order violation, and CAN tunnel message injection.
A comprehensive study of anomaly detection algorithms will be part of future work.

We compare two algorithms for each scenario.
Each scenario is tested with an Autoencoder with two hidden ReLu layers (size 32) and a ReLu code layer (size 4) that is trained (3 epochs, batch size 1, with shuffle) on the according baseline streams for every scenario.
One alternative anomaly detection algorithm is trained per scenario.
In future work, comparison with other existing detection tools is possible.

Used input metrics are bandwidth, avg.\ frame size, avg.\ frame gap, and avg.\ cycle jitter over \SI{100}{\milli\second} intervals.
This leads to up to \SI{1200} data points for training and up to \SI{120} data points for testing.
Total number of \ac{NADS} observation intervals is slightly lower as they are triggered by incoming packets that may not arrive precisely at the end of an interval. 
Thus, the interval is slightly longer dependent on the last incoming packet. 
Metrics calculations operate on real interval lengths.

Following scenarios each observe one stream with one type of anomaly applied.
Results show \acp{TN}, \acp{TP}, \acp{FN}, \acp{FP}, and precision and recall scores.

\begin{table}[b]
    \caption{In-car network streams with their priorities (PCP).\hspace{\textwidth}[zC = zonalController, * = wildcard]}
    \label{tab:sim_traffic}
    \centering
	\renewcommand{\arraystretch}{0.9}
	\setlength{\tabcolsep}{2.7pt}
    \begin{tabularx}{\columnwidth}{ccYc}
		\toprule
		\textbf{PCP} & \textbf{Sources} & \textbf{Destinations} & \textbf{Streams}\\
		\midrule
		7 & masterClock & * & 1x Time sync \\
		2 & connectivityGateway & adas & 1x incoming V2X data \\ 
		\cmidrule(lr){1-4}
		\multicolumn{4}{c}{\textbf{Timed Control Traffic}} \\
		\cmidrule(lr){1-4}	
		6 & zCFrontLeft & zC* & 3x manual driving control \\
		6 & adas & zC* & 3x auto driving control \\
		\cmidrule(lr){1-4}
		\multicolumn{4}{c}{\textbf{Shaped Data Streams}} \\
		\cmidrule(lr){1-4}
		5 & camera* & adas & 2x camera video \\
		5 & lidar* & adas & 4x raw lidar data \\
		\cmidrule(lr){1-4}
		\multicolumn{4}{c}{\textbf{CAN Tunnels}} \\
		\cmidrule(lr){1-4}
		4 & zCFrontLeft & zC*, infotainment & 42x CAN signal\\
		4 & zCFrontRight & zC*, infotainment & 61x CAN signal\\
		4 & zCRearLeft & zC*, infotainment & ~7x CAN signal\\
		4 & zCRearRight & zC*, infotainment & 78x CAN signal\\	
		4 & infotainment & zC* & 14x CAN signal\\
		\midrule
		\multicolumn{4}{r}{\textbf{Total Number of Streams:}} 216 \\
		\bottomrule
	\end{tabularx}
\end{table}

Goal of our case study is to show the versatility of the framework to compare different configurations in different scenarios. 
Future work will search for promising candidates and may recommend \ac{NADS} configurations.

\subsection{Baseline Scenario Configuration}
\label{subsec:case_study_baseline}
\begin{figure}[h!]
	\centering
	\includegraphics[width=0.9\columnwidth, trim= 0.9cm 1.5cm 0.9cm 1.5cm, clip=true]{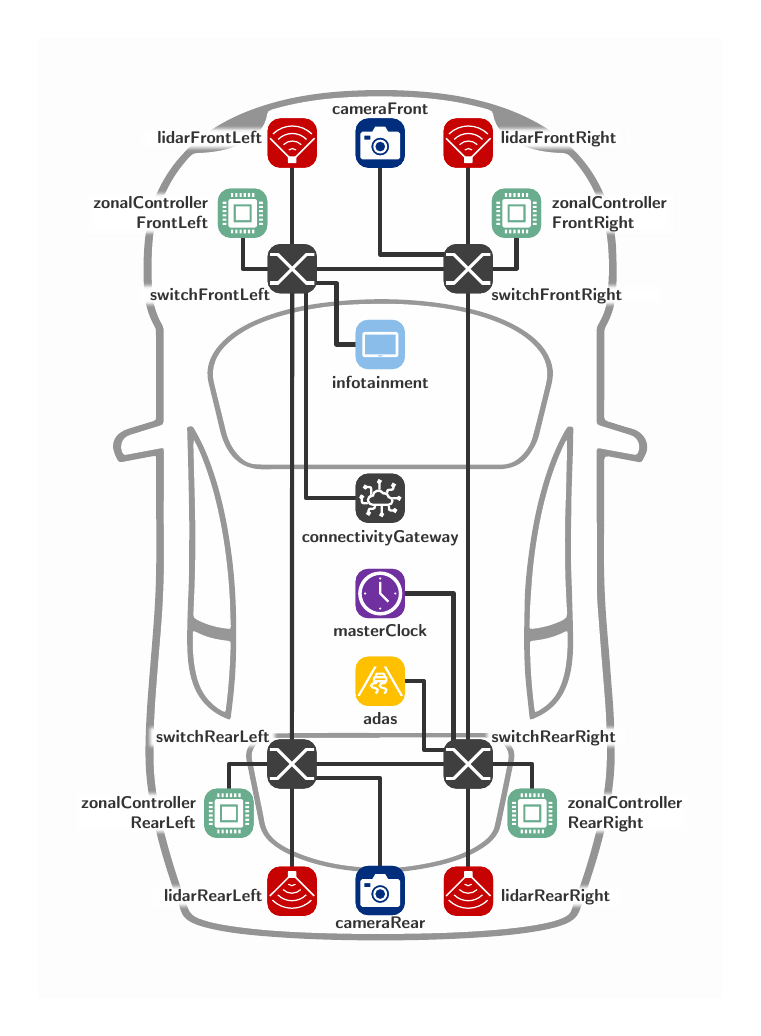}
	\caption{Sample \ac{IVN} topology with a zonal ring architecture.}
	\label{fig:sim_car_topology}
\end{figure}

The baseline scenario establishes a realistic in-car \ac{TSN} backbone.
We derive the \ac{IVN} topology from upcoming zonal designs with redundancy (s. Fig.~\ref{fig:sim_car_topology}).
All links operate at \SI{1}{\giga\bit\per\second}.
\ac{CAN} components derived from a production vehicle connect to four zonal controllers and the infotainment, serving as gateways~\cite{mhlsd-dsivi-20}. 
They exchange Ethernet-embedded \ac{CAN} messages of \acp{ECU} located in their respective physical zones. 
Clock drift is simulated, and all participants synchronize their clocks to a master using gPTP.
The setup includes four LIDARs and two cameras at the front and rear. 
A connectivity gateway connects the car to the outside world, and a central \ac{ADAS} oversees sensor fusion and driving control.
The static traffic configuration is derived from real application traces and an OEM in-car communication matrix.
The configuration covers all communication in every driving situation.

For our case study, we generate training and test data.
Cycles of the packet sources are between \SI{29}{\micro\second} and \SI{2}{\second}, with a mean packet interval of \SI{286.598}{\milli\second}.
One baseline simulation takes about \SI{3}{hours} on current consumer hardware and generates up to \SI{50}{\giga\byte} of \acp{PCAP}.
The resulting baseline dataset for training contains \acp{PCAP}-traces of \SI{120}{\second} (up to \SI{19027513} packets per switch).
Test data is created by applying an abnormal configuration to the baseline.
Abnormal simulation runs cover a shorter time span of \SI{12}{\second} (up to \SI{2002308} packets per switch).

Table \ref{tab:sim_traffic} shows the traffic streams with their priority (\ac{PCP}) matching three \ac{TSN} shaping configurations.
Timed control traffic (\ac{PCP} 6) uses a \ac{TDMA} scheme based on \ac{TAS}.
Packets are forwarded in exclusive time slots with a per switch offset to allow immediate forwarding.
The shaped data streams (\ac{PCP} 5) utilize \ac{CBS} to shape the traffic to a reserved bandwidth per link.
The \ac{ADAS} receives a maximum of \SI{900}{\mega\bit\per\second} raw video and LIDAR data.
The traffic patterns are derived from real LIDARs and cameras.
Furthermore, all timed and shaped streams (\ac{PCP} 6 and 5) use redundant paths through the ring topology.
The first switch forwards the streams in both directions and duplicate frames are eliminated for time control data at the zonal controllers, and for camera and LIDAR streams at the final switch connected to the \ac{ADAS}.
Streams that tunnel \ac{CAN} messages follow an OEM communication matrix.
The original \acp{ECU} are assigned to the physical zones, and messages are sent via the backbone for receiving \acp{ECU} that are not located in the zone of the sender.
Furthermore, gPTP traffic traverses with the highest priority (\ac{PCP} 7) between all devices to synchronize the clocks.
A TCP connection handles \ac{V2X} communication between the \ac{ADAS} and a connectivity gateway.
Traffic captures are recorded for incoming packets on all switch interfaces.

\subsection{Validation of the Baseline Scenario}
\label{subsec:baseline_validation}
A mandatory step in simulation-based assessments is the validation of the produced results to ensure that no errors or misconfigurations distort the simulation.
This section shows a selection of key validation results for the baseline scenario.

If all applications work as expected, the sent and received packet counts should be almost equal. 
A minor difference is acceptable due to packets still in transit when the simulation is stopped. 
There is no point in time when no packets are in flight.
In our case, the maximum difference between sent and received packets is 1, confirming that there is no packet loss.

\begin{figure}[h!]
	\begin{tikzpicture}
		\pgfplotsset{every axis/.style={
				width=1\columnwidth,
				height=0.5\columnwidth,
				ylabel = End-to-end latency [\si{\micro\second}],
				ybar,
				ymin = 0,
				ymax = 0.000100,
				change y base,
      			y SI prefix=micro,
				bar width = 4pt,
				xtick = data,
				table/col sep = comma,
				xtick pos=bottom,
				x tick label style = {rotate=20, anchor=east},
				xticklabels = {auto brake, auto steer, auto throttle, manual brake, manual steer, manual throttle},
				legend columns=5,
				legend  style={at={(0.5 ,1.36)}, legend cell align=left, anchor=north, font=\footnotesize},
		}}
		\def \pathToCsv {data/General_latencies_prio6_streams.csv}
		\begin{axis}[bar shift=-.3cm, hide axis]
			\addplot[plotred, fill, bar width=3pt] table[x expr =\coordindex , y = zonalControllerRearRightMaxL , col sep = comma] {\pathToCsv}; \label{plot_zonalControllerRearRight_max}
			\addplot[plotorange, fill] table[x expr =\coordindex , y = zonalControllerRearRightMinL , col sep = comma] {\pathToCsv}; \label{plot_zonalControllerRearRight_min}
		\end{axis}
		\begin{axis}[bar shift=-.1cm, hide axis]
			\addplot[plotred, fill, bar width=3pt] table[x expr =\coordindex , y = zonalControllerRearLeftMaxL , col sep = comma] {\pathToCsv}; \label{plot_zonalControllerRearLeft_max}
			\addplot[plotyellow, fill] table[x expr =\coordindex , y = zonalControllerRearLeftMinL , col sep = comma] {\pathToCsv}; \label{plot_zonalControllerRearLeft_min}
		\end{axis}
		\begin{axis}[bar shift=.1cm, hide axis]
			\addplot[plotred, fill, bar width=3pt] table[x expr =\coordindex , y = zonalControllerFrontRightMaxL , col sep = comma] {\pathToCsv}; \label{plot_zonalControllerFrontRight_max}
			\addplot[plotgreen, fill] table[x expr =\coordindex , y = zonalControllerFrontRightMinL , col sep = comma] {\pathToCsv}; \label{plot_zonalControllerFrontRight_min}
		\end{axis}
		\begin{axis} [bar shift=.3cm]
			\addlegendimage{/pgfplots/refstyle=plot_zonalControllerRearRight_max}\addlegendentry{\textbf{Jitter} (=}
			\addlegendimage{empty legend}
			\addlegendentry{\hspace{-.6cm}Max - Min)}
			\addlegendimage{empty legend}
			\addlegendentry{\hspace{-.25cm}\textbf{Latency}}
			\addlegendimage{empty legend}
			\addlegendentry{\hspace{-.6cm}zonalController...}
			\addlegendimage{empty legend}
			\addlegendentry{\hspace{-.6cm}}
			\addlegendimage{/pgfplots/refstyle=plot_zonalControllerRearRight_min}\addlegendentry{...RearRight}
			\addlegendimage{/pgfplots/refstyle=plot_zonalControllerRearLeft_min}\addlegendentry{...RearLeft}
			\addlegendimage{/pgfplots/refstyle=plot_zonalControllerFrontRight_min}\addlegendentry{...FrontRight}
			\addlegendimage{/pgfplots/refstyle=plot_zonalControllerFrontLeft_min}\addlegendentry{...FrontLeft}
			\addplot[plotred, fill, bar width=3pt] table[x expr =\coordindex , y = zonalControllerFrontLeftMaxL , col sep = comma] {\pathToCsv}; \label{plot_zonalControllerFrontLeft_max}
			\addplot[plotblue, fill] table[x expr =\coordindex , y = zonalControllerFrontLeftMinL , col sep = comma] {\pathToCsv}; \label{plot_zonalControllerFrontLeft_min}
		\end{axis}
	\end{tikzpicture}
	\vspace{-16pt}
	\caption{Minimum and maximum end-to-end latency of the timed control traffic per receiver (cf. Table \ref{tab:sim_traffic}).}
	\label{fig:baseline_latencies_pcp6}
	\vspace{-10pt}
\end{figure}
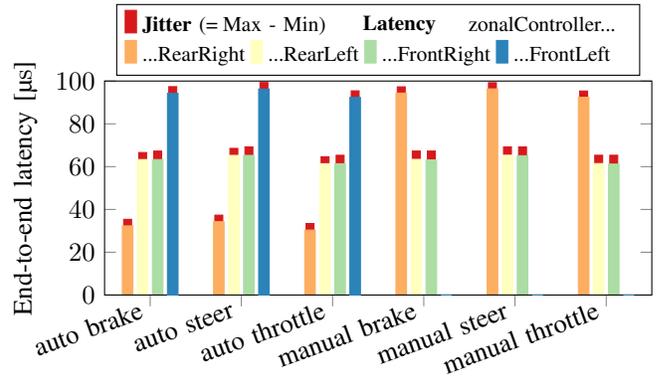

The end-to-end latencies can help to validate the timing of the network.
Figure \ref{fig:baseline_latencies_pcp6} shows the max. and min. end-to-end latencies of the six timed control streams (\ac{PCP} 6) for each receiver.
The latency depends on the hop count of one to four switches.
Each hop offsets the forwarding windows up to \SI{30}{\micro\second} after packet reception to compensate for possible clock synchronization inaccuracies.
The maximum latency of a packet is the sum of the forwarding windows of all switches and the transmission time of the packet: 
\SIrange{30}{37}{\micro\second} for one hop, \SIrange{61}{69}{\micro\second} for two hops, and \SIrange{92}{99}{\micro\second} for three hops.
Four hops are never used because there is no packet loss and the redundant path is never providing the received packet.
Jitter is the difference between the maximum and minimum latency of a stream.
In the scenario, max. per stream jitter is \SI{4}{\micro\second}, aligning with synchronization accuracy expectations for timed control traffic streams in the low microsecond range.

The maximum latency of lower priorities is \SI{157}{\micro\second} for the shaped data streams (\ac{PCP} 5), \SI{268}{\micro\second} for the \ac{CAN} tunnel streams (\ac{PCP} 4), and \SI{224}{\micro\second} for the \ac{V2X} stream (\ac{PCP} 2).

The validation excerpt indicates that the baseline scenario is configured correctly and produces packet traces as intended. 
Following abnormal scenarios change characteristics of this baseline and are validated in their respective sections.

\subsection{Eliminate Packets of the Auto Brake Stream}
\label{subsec:eliminate_auto_brake}

Auto brake is a \ac{TAS} auto driving control stream (cf. Tab. \ref{tab:sim_traffic}) that sends frames with a size of \SI{110}{\byte} via multicast in a cycle of \SI{1}{\milli\second} from the \ac{ADAS} to all zonal controllers. 
The zonal controllers directly drive the brakes or forward them through an internal gateway onto a \ac{CAN} bus.

Elimination takes place in switch rear right and affects all packets leaving in counterclockwise direction.
The scenario alternates between elimination for \SI{1}{\second} and no elimination for \SI{1}{\second}.
To generate a sporadic low-threshold anomaly, a packet is eliminated with a probability of \SI{50}{\percent} and a minimum clearance between eliminations of \SI{10}{\milli\second} in the active phases.

Real-world consequences could range from an extended braking distance to \ac{ESC} malfunction or a complete loss of braking functionality.

In the generated traces the scenario is visible through missing packets.
However, the simulation results show that there is still no packet loss at the receiver.
This is since the packets now arrive via the longer redundant 4-hop path.
As a result, maximum latency of the auto brake stream increases to \SI{128}{\micro\second} with a jitter of \SI{65}{\micro\second}.

The \ac{NADS} is trained with the baseline scenario \ac{PCAP} filtered for the auto brake stream on the incoming port of switch front right.
Tests are performed with scenario specific \acp{PCAP} using the same filter.

This scenario compares Autoencoder and mean-shift based algorithms.
The size of created clusters is used as the border of normal and abnormal inputs.
After learning, inputs are considered normal when within any cluster and abnormal when outside all clusters.
A scale factor can be configured to adapt the cluster sizes, which we set to 1.1 in this scenario.

Table \ref{tab:eliminate_auto_brake} shows the detection scores of an Autoencoder and a mean-shift based \ac{NADS} in the elimination scenario.
The mean-shift algorithm performs perfectly (precision and recall value 1.00) within this scenario and time span.
The Autoencoder detection performs with a precision of 0.90 and a recall of 1.00.
Results show six \ac{FP} in the test set.

\begin{table}[h!]
	\caption{Elimination of auto brake stream packets -- Scores of a \ac{NADS} based upon Autoencoder and Mean-shift clustering.}
	\label{tab:eliminate_auto_brake}
	\centering
	\renewcommand{\arraystretch}{0.9}
	\setlength{\tabcolsep}{2.7pt}
	\begin{tabularx}{\columnwidth}{ccYYYY}
		\toprule		                                             
		\multicolumn{2}{c}{\textbf{Detection Algorithm}} & \multicolumn{2}{c}{Autoencoder}                   & \multicolumn{2}{c}{Mean-shift} \\
		\cmidrule(lr){1-6}
		\multicolumn{2}{c}{\textbf{Predicted Class}}     & Abnormal                & Benign                  & Abnormal                & Benign \\	
		\cmidrule(lr){1-2}	                             \cmidrule(lr){3-4}                                  \cmidrule(lr){5-6}
		\multirow{2}{*}{\textbf{True Class}} & Abnormal  & \colorbox{green!30}{55} TP & ~\colorbox{red!30}{0} FN   & \colorbox{green!30}{55} TP & ~\colorbox{red!30}{0} FN \\
						                     & Benign    & ~\colorbox{red!30}{6} FP   & \colorbox{green!30}{47} TN & ~\colorbox{red!30}{0} FP   & \colorbox{green!30}{53} TN \\
		\cmidrule(lr){1-2}							     \cmidrule(lr){3-4}                                  \cmidrule(lr){5-6}
        \multicolumn{2}{c}{\textbf{Precision} = TP / (TP + FP)}           & \multicolumn{2}{c}{\ApplyGradient{0.90}}          & \multicolumn{2}{c}{\ApplyGradient{1.00}} \\
		\multicolumn{2}{c}{~~~~\textbf{Recall} = TP / (TP + FN)}              & \multicolumn{2}{c}{\ApplyGradient{1.00}}          & \multicolumn{2}{c}{\ApplyGradient{1.00}} \\
		\bottomrule
	\end{tabularx}
\end{table}

But even those high scores are not sufficient for an independent real-world deployment.
Even one \ac{FP} every \SI{12}{\second} that directly lead to cockpit warnings or countermeasures would be unacceptable.
In-car \ac{NADS} need to be filtered, combined, and tuned to reduce \acp{FP} to an acceptable minimum.

\subsection{Reorder Packets of the Camera Front Stream}
\label{subsec:reorder_camera_front}

Camera front stream is shaped by \ac{CBS} (cf. Tab. \ref{tab:sim_traffic}), send frames have a size of \SI{1426}{\byte} and are uniformly distributed in a cycle of \SIrange{30}{100}{\micro\second} (approx. \SI{176}{\mega\bit\per\second}).

Reordering takes place in switch front right and affects all packets leaving in clockwise direction.
The reordering scenario is alternating active for \SI{1}{\second} where a packet is reordered behind the next packet with a probability of \SI{50}{\percent} and a minimum clearance of \SI{10}{\milli\second} and inactive for \SI{1}{\second}.

Violations of order can lead to consequences such as packet loss, or unspecified behavior in the target application layer stack. 
For video streams this can lead to loss of frames or a loss of synchronization between video and other sensor data.

Generated \acp{PCAP} show missing frames that are taken and injected after following video stream.
All packets still arrive at the destination.
The maximum latency of the camera front stream increases from \SI{121}{\micro\second} in the baseline scenario to \SI{181}{\micro\second}.

The \ac{NADS} is trained with the baseline \ac{PCAP} filtered for the stream at which transports raw data from camera front.
The filtered interface is the incoming port in switch rear right.
Again, the tests use \acp{PCAP} recorded in the according scenario and filtered the same way.

Table \ref{tab:reorder_camera_front} shows the detection scores in the reorder scenario for an Autoencoder and a \ac{NADS} based on the Isolation Forest algorithm with automatic maximum samples and a contamination of 0.1.
In case of the Autoencoder, the number of false predictions is high with 8 \acp{FP} and 35 \acp{FN}.
With a precision of 0.71 and a recall of 0.36 the detection scores are low compared to the isolation forest with just 4 \acp{FP} and almost all anomalies detected (recall of 0.96).

\begin{table}[h!]
	\caption{Reordering of front camera stream packets -- Scores of a \ac{NADS} based upon Autoencoder and Isolation Forest.}
	\label{tab:reorder_camera_front}
	\centering
	\renewcommand{\arraystretch}{0.9}
	\setlength{\tabcolsep}{2.7pt}
	\begin{tabularx}{\columnwidth}{ccYYYY}
		\toprule		                                             
		\multicolumn{2}{c}{\textbf{Detection Algorithm}} & \multicolumn{2}{c}{Autoencoder}                   & \multicolumn{2}{c}{Isolation Forest} \\
		\cmidrule(lr){1-6}
		\multicolumn{2}{c}{\textbf{Predicted Class}}     & Abnormal                & Benign                  & Abnormal                & Benign \\	
		\cmidrule(lr){1-2}	                             \cmidrule(lr){3-4}                                  \cmidrule(lr){5-6}
		\multirow{2}{*}{\textbf{True Class}} & Abnormal  & \colorbox{green!30}{20} TP & \colorbox{red!30}{35} FN   & \colorbox{green!30}{53} TP & ~\colorbox{red!30}{2} FN \\
						                     & Benign    & ~\colorbox{red!30}{8} FP   & \colorbox{green!30}{45} TN & ~\colorbox{red!30}{4} FP   & \colorbox{green!30}{49} TN \\
		\cmidrule(lr){1-2}							     \cmidrule(lr){3-4}                                  \cmidrule(lr){5-6}
        \multicolumn{2}{c}{\textbf{Precision} = TP / (TP + FP)}           & \multicolumn{2}{c}{\ApplyGradient{0.71}}          & \multicolumn{2}{c}{\ApplyGradient{0.93}} \\
		\multicolumn{2}{c}{~~~~\textbf{Recall} = TP / (TP + FN)}              & \multicolumn{2}{c}{\ApplyGradient{0.36}}          & \multicolumn{2}{c}{\ApplyGradient{0.96}} \\
		\bottomrule
	\end{tabularx}
\end{table}

\subsection{Inject Packets into a CAN Tunnel Stream}
\label{subsec:inject_can_tunnel}

A \ac{CAN} tunnel stream is a legacy control signal to which strict priority shaping is applied (cf. Tab. \ref{tab:sim_traffic}).
This scenario considers a tunnel which transports a \ac{CAN} payload of \SI{8}{\byte} as UDP payload in a fixed cycle of \SI{60}{\milli\second}.
The observed flow in this scenario starts at zonal controller rear left and is forwarded to the controller front left and front right.

Injection takes place at switch rear left, with all packets leaving in counterclockwise direction.
The injection scenario is alternating active for \SI{1}{\second} and inactive for \SI{1}{\second}.
In active phases a packet with \SI{8}{\byte} payload and identical header values is injected every \SI{50}{\milli\second} with a probability of \SI{50}{\percent}.

The injection of packets into \ac{CAN} is a common attack vector on traditional \ac{CAN} buses. 
It has been used in real production cars to alter the behavior of diverse functions~\cite{mv-reupv-15}.

The generated traces contain normal and injected packets.
Simulation results show that the receiver of the \ac{CAN} stream is receiving 41 packets more than sent from source.
The \ac{NADS} is trained with the baseline \ac{PCAP} filtered for the single observed \ac{CAN} tunnel stream on the incoming port in switch front left.
The same filter is applied to the injection scenario \acp{PCAP}.

In this case, the Autoencoder is compared with a \ac{HBOS} with 10 bins and a contamination of 0.1.
Table \ref{tab:inject_can_tunnel} shows the detection scores of an Autoencoder and a \ac{HBOS} based \ac{NADS} in the injection scenario.
The number of packets (cycle of \SI{60}{\milli\second}) in a monitored interval (\SI{100}{\milli\second}) is at minimum (2).
Even with minimum number of packets per observation interval the Autoencoder performs with a precision of 0.76 and a recall of 0.89.
\acp{FP} indicate that the traffic pattern is learned to some extent and the share of detected injections is high.
The \ac{HBOS} based \ac{NADS} performs with lower precision of 0.74 while maintaining perfect recall.

\begin{table}[h!]
\caption{Injection of packets into a \ac{CAN} tunnel stream -- Scores of a \ac{NADS} based upon Autoencoder and HBOS.}
\label{tab:inject_can_tunnel}
\centering
\renewcommand{\arraystretch}{0.9}
\setlength{\tabcolsep}{2.7pt}
\begin{tabularx}{\columnwidth}{ccYYYY}
	\toprule		                                             
	\multicolumn{2}{c}{\textbf{Detection Algorithm}} & \multicolumn{2}{c}{Autoencoder}                   & \multicolumn{2}{c}{HBOS} \\
	\cmidrule(lr){1-6}
	\multicolumn{2}{c}{\textbf{Predicted Class}}     & Abnormal                & Benign                  & Abnormal                & Benign \\	
	\cmidrule(lr){1-2}	                             \cmidrule(lr){3-4}                                  \cmidrule(lr){5-6}
	\multirow{2}{*}{\textbf{True Class}} & Abnormal  & \colorbox{green!30}{25} TP & ~\colorbox{red!30}{3} FN   & \colorbox{green!30}{28} TP & ~\colorbox{red!30}{0} FN \\
										 & Benign    & ~\colorbox{red!30}{8} FP  & \colorbox{green!30}{53} TN & ~\colorbox{red!30}{10} FP   & \colorbox{green!30}{51} TN \\
	\cmidrule(lr){1-2}							     \cmidrule(lr){3-4}                                  \cmidrule(lr){5-6}
	\multicolumn{2}{c}{\textbf{Precision} = TP / (TP + FP)}           & \multicolumn{2}{c}{\ApplyGradient{0.76}}          & \multicolumn{2}{c}{\ApplyGradient{0.74}} \\
	\multicolumn{2}{c}{~~~~\textbf{Recall} = TP / (TP + FN)}              & \multicolumn{2}{c}{\ApplyGradient{0.89}}          & \multicolumn{2}{c}{\ApplyGradient{1.00}} \\
	\bottomrule
\end{tabularx}
\end{table}

\subsection{Findings}
\label{subsec:findings}

The presented scenarios demonstrate the potential of our approach by adjusting three options: revealing how traffic patterns, anomaly types, and detection algorithms impact detection performance.
Rapid checks of different algorithms and configurations are possible. 
Longer runs enable comprehensive assessments of numerous scenarios with increased number of data points and upscaled systematic \ac{NADS} parameter studies.

The shown anomalies are minimal deviations from baseline traffic pattern.
Our scenarios show a small subset of variations that can be performed.
More sophisticated scenarios are needed for extensive assessment and candidate selection.

Rapid assessment of \ac{NADS} performance is essential for candidate selection.
Parallelization minimizes simulation time for multiple scenarios, as each simulation runs in a separate process.
Scaling is constrained only by available hardware.
Furthermore, a dataset generated from one simulation can be reused for assessments across all streams and various \acp{NADS}.

\section{Conclusion and Outlook}
\label{sec:conclusion}

\acp{NADS} are essential for securing future automotive systems and allow for mitigating risks by implementing countermeasures on misbehavior.  
Diverse \ac{NADS} technologies are available and the selection of suitable candidates is a complex task.
Often the assessment of \ac{NADS} candidates relies on comparing its performance on characteristic datasets, which are not easily available nor suitable to represent domain-specific features.

We presented an assessment framework for \acp{NADS}, which is based on system level network simulation.
Our tools are adaptable and extendable and can incorporate real-world data.
Our dataset generation allows for the creation of events that are tailored to domain-specific needs while maintaining a widespread and standardized format.

Modular and extensible \ac{NADS} implementations allow evaluating different technologies and comparing them with each other.
Detection scores showed that \ac{NADS} performance depends on \ac{TSN} traffic class, anomaly type, and \ac{NADS} implementation details.
Our assessment toolchain proved suitable for a fast and insightful examinations.

This work opens the following future research directions.

(1) Systematic studies in complex scenarios under various \ac{NADS} configurations.
A large-scale comparison of \ac{NADS} methods in the \ac{IVN} domain can show which combinations of traffic class and detection method work effectively.

(2) Integration of realistic in-car attacks and their intrusion data footprint.
Our toolchain allows for incorporating sophisticated attack models within its datasets.
A catalog of attack models can be created to provide a set of realistic attacks.

(3) Parameter tuning of \ac{NADS} methods.
With our tooling the evaluation of \ac{NADS} methods with different parameter settings is possible.
It can be used to automate the parameter tuning process and to find optimal parameter settings.

(4) Evaluation of \ac{NADS} methods in further hard real-time Ethernet domains.
The toolchain is transferable to hard real-time Ethernet domains such as industrial facilities or airplanes.

\bibliographystyle{IEEEtran}
\bibliography{bibtex/HTML-Export/all_generated,bibl/special,local}

\end{document}